\input{aipcheck}

\documentclass[
    ,final            
  ]
  {aipproc}

\layoutstyle{6x9}

\newcommand{\ba}{\begin{eqnarray}}
\newcommand{\ea}{\end{eqnarray}}
\newcommand{\ban}{\begin{eqnarray*}}
\newcommand{\ean}{\end{eqnarray*}}
\newcommand{\bsub}{\begin{subequations}}
\newcommand{\esub}{\end{subequations}}

\newcommand{\nc}{\newcommand}
\nc{\Id}{{\mathchoice {\rm 1\mskip-4mu l} {\rm 1\mskip-4mu l}
{\rm 1\mskip-4.5mu l} {\rm 1\mskip-5mu l}}}

\begin{document}

\title{Symmetries at and Near Critical Points of Quantum
Phase Transitions in Nuclei}

\author{A. Leviatan}{
  address={Racah Institute of Physics, The Hebrew University, 
Jerusalem 91904, Israel}
}

\author{F. Iachello}{
  address={Center for Theoretical Physics, 
Sloane Physics Laboratory, 
Yale University, P.O. Box 208120, New Haven, CT 06520-8120, USA}
}

\begin{abstract}
We examine several types of symmetries which are 
relevant to 
quantum phase transitions in nuclei. 
These include: critical-point, quasidynamical, and 
partial dynamical symmetries.  

\end{abstract}
 
\maketitle


Symmetry plays a profound role in thermal and quantum phase transitions 
(QPT). The latter occur at zero temperature as a function of a 
coupling constant in the Hamiltonian. 
Such ground-state energy phase transitions are a pervasive 
phenomenon observed in many branches of physics, and are realized 
empirically in nuclei as transitions between different shapes. 
QPT occur as a result of a competition between terms in the Hamiltonian 
with different symmetry character, which lead to considerable mixing in the 
eigenfunctions, especially at the critical-point where the structure 
changes most rapidly. In the present contribution we address the question 
whether there are any symmetries (or traces of) still present at and near 
critical points of QPT. As shown below, unexpectedly, several types of 
intermediate-symmetries can survive in spite of the strong mixing.

The first indication that symmetries can persist at critical points of 
QPT came from the recent recognition that the dynamics at these 
critical-points is amenable to analytic descriptions~\cite{iac00,iac01}. 
For nuclei, such analytic benchmarks of criticality, called 
``critical-point symmetries'', were obtained in the geometric framework 
of a Bohr Hamiltonian for macroscopic quadrupole shapes. 
The E(5) benchmark \cite{iac00}
is applicable to a second-order shape-phase transition between
spherical and deformed $\gamma$-soft nuclei. 
The X(5) benchmark \cite{iac01} is applicable to a low-barrier 
first-order phase transition between spherical and axially-deformed nuclei. 
Both benchmarks employ an infinite square-well potential,  
which is $\gamma$-independent for $E(5)$ and axially-deformed, with an 
assumed $\beta\gamma$ decoupling, for $X(5)$. The eigenvalues are 
proportional to squared zeroes of Bessel functions, and display 
spectral features which are in-between the indicated limiting phases. 
The importance of these analytic benchmarks lies 
in the fact that they provide a classification of states (quantum numbers) 
and analytic expressions (parameter-free except for scale) for observables 
in regions of rapid structural changes. Empirical evidence for these 
``critical-point symmetries'' 
has been presented in nuclei~\cite{casten00,casten01}. 
An example of $X(5)$-like structure found in $^{152}$Sm and $^{150}$Nd 
is shown in Fig.~1.
\begin{figure}[t]
\hspace{-0.13cm}
\includegraphics[height=.3\textheight]{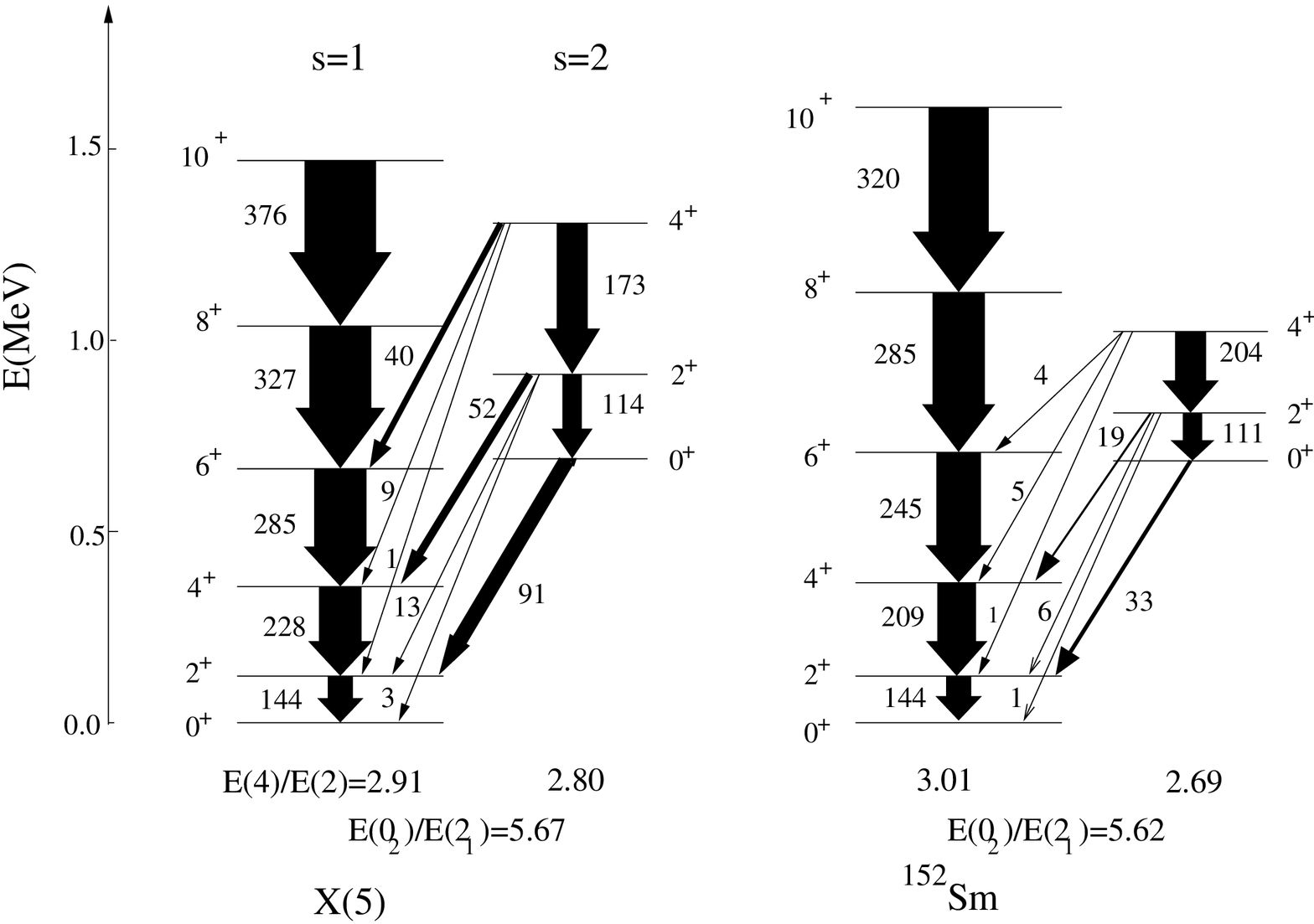}
  \includegraphics[height=.25\textheight,angle=90]{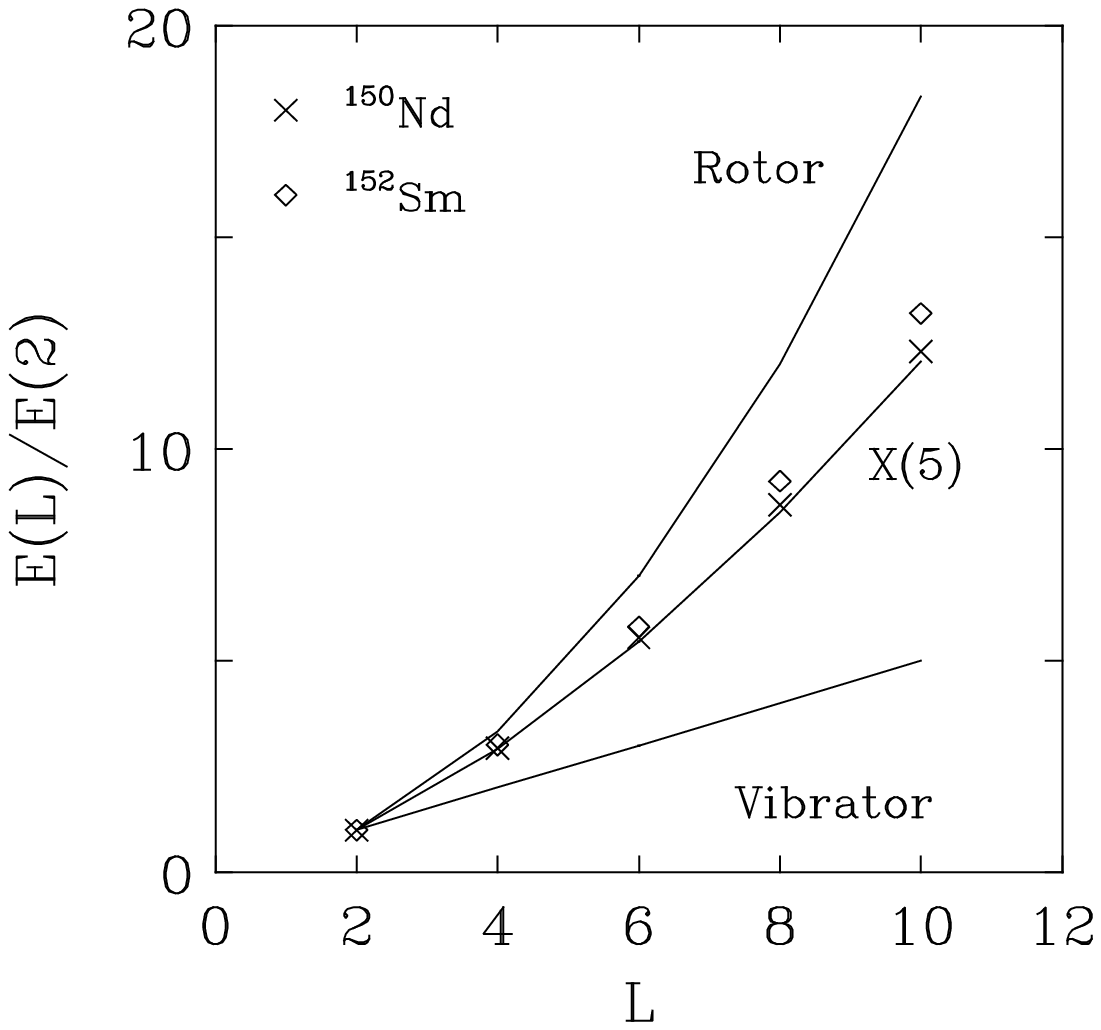}
  \caption{Energy spectrum of $^{152}$Sm and $^{150}$Nd 
compared to that of the X(5) ``critical-point symmetry''. From 
~\cite{casten01}.}
\vspace{-1cm}
\end{figure}

A convenient framework to study symmetry-aspects of QPT 
is the interacting boson model (IBM)~\cite{ibm}, 
which describes quadrupole collective states in 
nuclei in terms 
of a system of $N$ monopole ($s$) and
quadrupole ($d$) bosons, representing valence nucleon pairs. 
The model is based on a $U(6)$ spectrum generating algebra.
Dynamical symmetries occur when the Hamiltonian
is written in terms of the Casimir operators 
of a chain of nested algebras of $U(6)$.
They provide analytically solvable limits and quantum 
numbers, which are the labels of the irreducible representations (irreps) 
of the algebras in the chain. The three dynamical symmetry limits of 
the IBM and associated bases are
\ba
&&U(6) \supset U(5)  \supset O(5) \supset O(3)
\qquad 
\vert N,n_d,\tau,\tilde{\nu},L\rangle \quad\;\;\;\,
{\rm spherical\; vibrator}\qquad\nonumber\\
&&U(6) \supset SU(3) \supset O(3)
\qquad\qquad\;\;\;
\vert N,(\lambda,\mu),K,L\rangle \quad
{\rm axially\; deformed\; rotor}\qquad\\
&&U(6) \supset O(6)  \supset O(5) \supset O(3)
\qquad
\vert N,\sigma,\tau,\tilde{\nu},L\rangle 
\qquad
\gamma{\rm -soft\;deformed\;  rotor}\qquad\nonumber
\label{o6ds}
\ea
The corresponding analytic solutions resemble spectral features 
of known geometric paradigms, indicated above. 
This identification is consistent with the geometric visualization 
of the model in terms of a potential surface defined by 
the expectation value of the Hamiltonian in a coherent (intrinsic) 
state~\cite{gino80,diep80}. For one- and two-body interactions the 
surface has the form  $E(\beta,\gamma) = E_0 + 
N(N-1)f(\beta,\gamma)$, with  
$f(\beta,\gamma) = (1+\beta^2)^{-2}
\beta^2\left [ a - b\beta\cos 3\gamma + c\beta^2 \right ]$. 
The quadrupole shape parameters $(\beta,\gamma)$ at the global minimum
of $E(\beta,\gamma)$ define the equilibrium shape for a given Hamiltonian. 
Each dynamical symmetry corresponds to a possible phase of the system.
Phase transitions can be studied 
by IBM Hamiltonians of the form~\cite{diep80}, 
\ba 
H(\alpha) &=& (1-\alpha)\,H_{1} + \alpha\, H_{2} ~,
\ea
involving 
terms from different dynamical symmetry chains. 
The nature of the phase transition is 
dictated by the topology of the underlying surface. 
The energy surfaces 
at the critical-points of first- and second-order transitions 
have the form 
\ba
&&1^{st}\, {\rm order}:\;\;\;
f(\beta,\gamma=0) = 
c(1+\beta^2)^{-2}\beta^2\left ( \beta-\beta_0\right )^2 ~, 
\quad\nonumber\\
&&2^{nd}\, {\rm order}:\;\;\;
f(\beta,\gamma) =  c(1+\beta^2)^{-2}\beta^4 ~.\quad
\ea
\label{1st2nd}
As shown in Fig.~2, the first-order critical-surface has degenerate 
spherical and deformed minima at $\beta=0$ and $(\beta=\beta_0>0,\gamma=0)$. 
The position ($\beta_{+}$) and height ($h$) of the barrier are indicated 
in the caption. The second-order critical-surface is independent of 
$\gamma$ and is flat bottomed $(\sim \beta^4)$ for small $\beta$. 
By requiring the Hamiltonian $H(\alpha)$ of Eq.~(2) to have 
a critical energy-surface, one pins down the value of the control 
parameter at the critical-point $(\alpha=\alpha_c)$. The critical-point 
Hamiltonians, obtained in this manner, 
for the $U(5)$-$SU(3)$ and $U(5)$-$O(6)$ phase transitions 
are given by  
\ba
&&U(5)-SU(3):\quad
H(\alpha) = (1-\alpha)\,\hat{n}_d  
-\alpha\frac{1}{4N}\,Q\cdot Q\;\;\;\;\; 
\alpha_c = \frac{16N}{(34N-27)} ~,\nonumber\\ 
&&U(5)-O(6):\quad\;\;\;
H(\alpha) = (1-\alpha)\,\hat{n}_d  + \alpha\frac{1}{N}\hat{P}_{6}\qquad
\;\;\;
\alpha_c =\frac{N}{(2N-1)} ~.\qquad\quad\;\;\;
\label{Hcri}
\ea 
IBM Hamiltonians of this kind have been studied extensively~\cite{iaczam04}, 
concluding that the $U(5)$-$SU(3)$ transition is of first order, 
with an extremely low barrier, (corresponding to $\beta_0 = \sqrt{2}/4$ 
and $h\approx 10^{-3}$ in Fig.~2). The $U(5)$-$O(6)$ transition is 
found to be of second order. The corresponding critical-point Hamiltonians, 
Eq.~(\ref{Hcri}), exhibit $X(5)$- and $E(5)$-like spectrum, respectively, 
albeit finite-N modifications~\cite{iaczam04,levgin03,lev05}. 
\begin{figure}[t]
\includegraphics[height=0.3\textheight,width=0.43\textwidth,
angle=270]{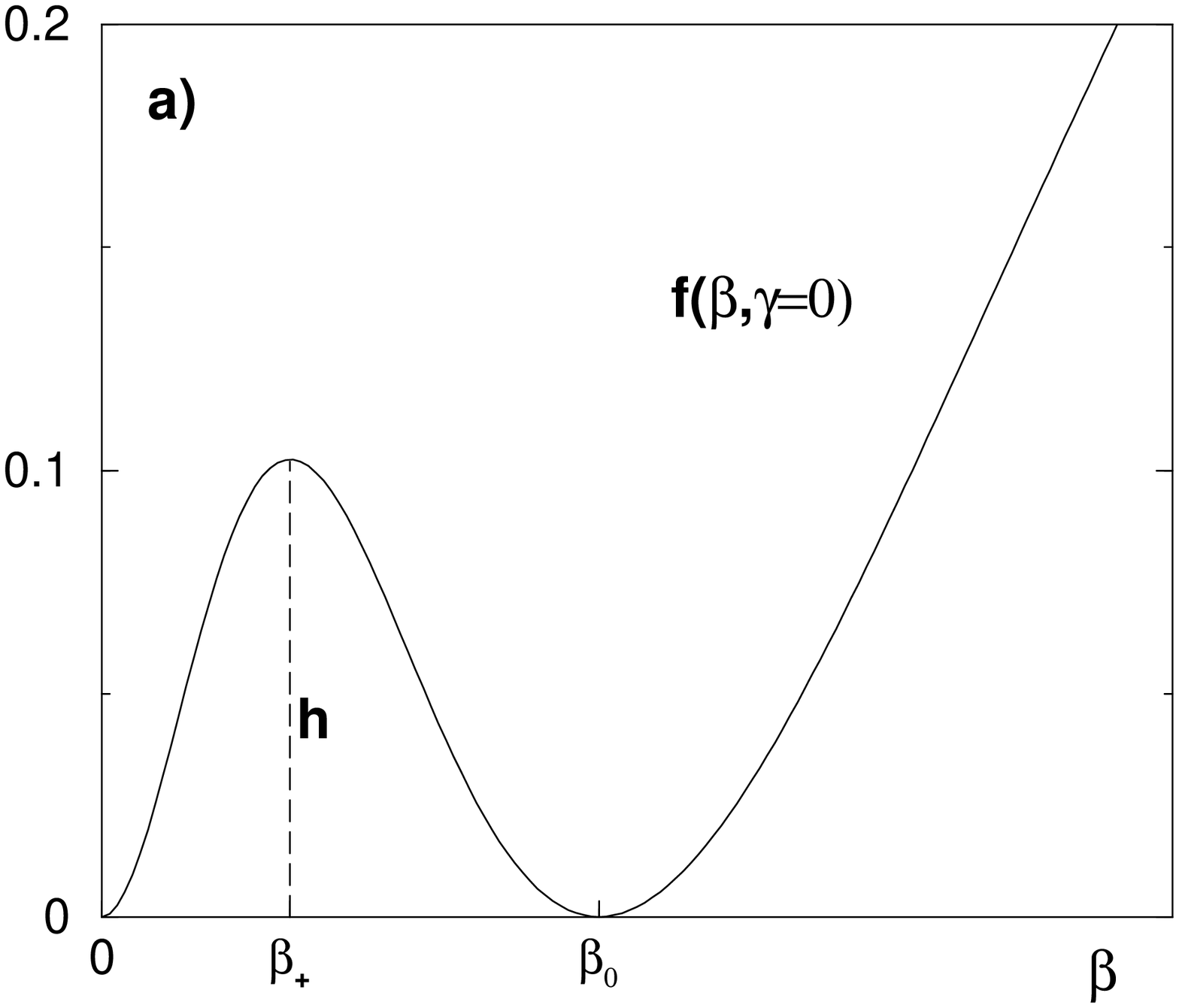}
\hspace{0.5cm}
  \includegraphics[height=0.3\textheight,width=0.43\textwidth,
angle=270]{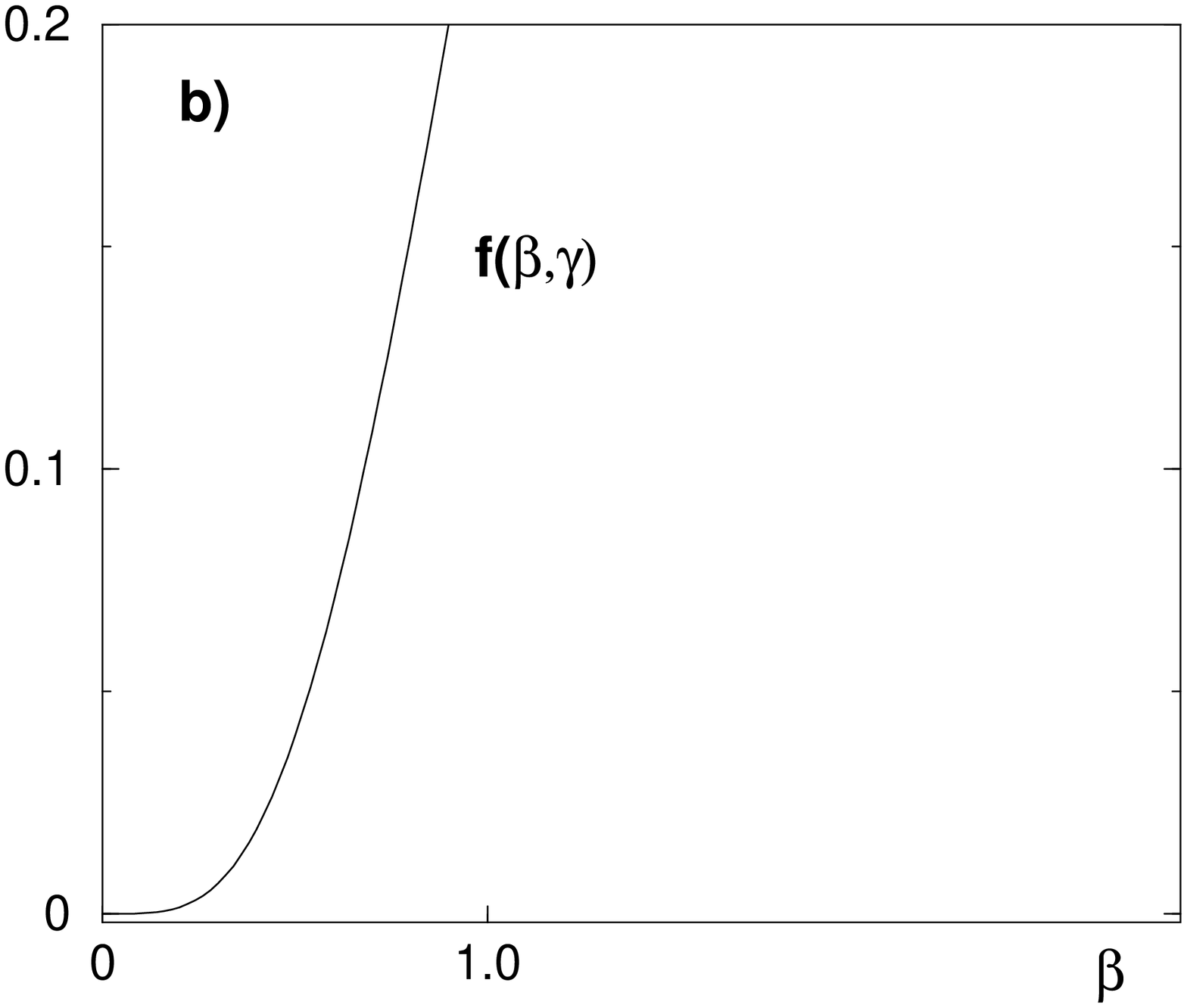}
  \caption{Energy surfaces at the critical points, Eq.~(3). 
(a) First-order transition. The position and height of the barrier 
are $\beta= \beta_{+} = (-1 + \sqrt{1+\beta_{0}^2}\;)/\beta_0\,$ 
and $h = f(\beta_{+}) = 
( -1 + \sqrt{1+\beta_{0}^2}\,)^2/4 \,$ respectively. 
(b) Second-order transition. In this case 
$f(\beta,\gamma)$ is independent of $\gamma$.}
\end{figure} 

From the point of view of symmetry, $H(\alpha)$, Eq.~(2),  
involves competing incompatible (non-commuting) symmetries. 
For $\alpha=0$ or $\alpha=1$, one recovers the limiting symmetries. 
For $0<\alpha<1$, both symmetries are broken and the mixing is particularly 
strong at the critical-point ($\alpha_c\approx1/2$). A detailed study of 
the symmetry content of the IBM Hamiltonians, Eq.~(4), 
upon variation of the control parameter $\alpha$, 
has found that for most values of $\alpha$, except for a narrow 
region near the critical-point $(\alpha=\alpha_c)$, 
selected low-lying states continue to exhibit characteristic properties 
(energy and B(E2) ratios) of the {\it closest} dynamical symmetry limit. 
Such an ``apparent'' persistence of symmetry in the face of strong 
symmetry-breaking interactions, was called  
``quasidynamical symmetry''~\cite{rowe04,rowe05}. The indicated 
persistence is clearly evident in the spectrum shown in Fig.~3, 
for the $U(5)$-$SU(3)$ transition. This ``apparent'' symmetry is due to the 
coherent nature of the mixing. 
As seen on the right hand side of Fig.~3, 
the mixing of $SU(3)$ irreps is large, but is 
approximately independent of the angular momentum of 
the yrast states.
\begin{figure}[t]
\includegraphics[height=.3\textheight,width=0.48\textwidth]{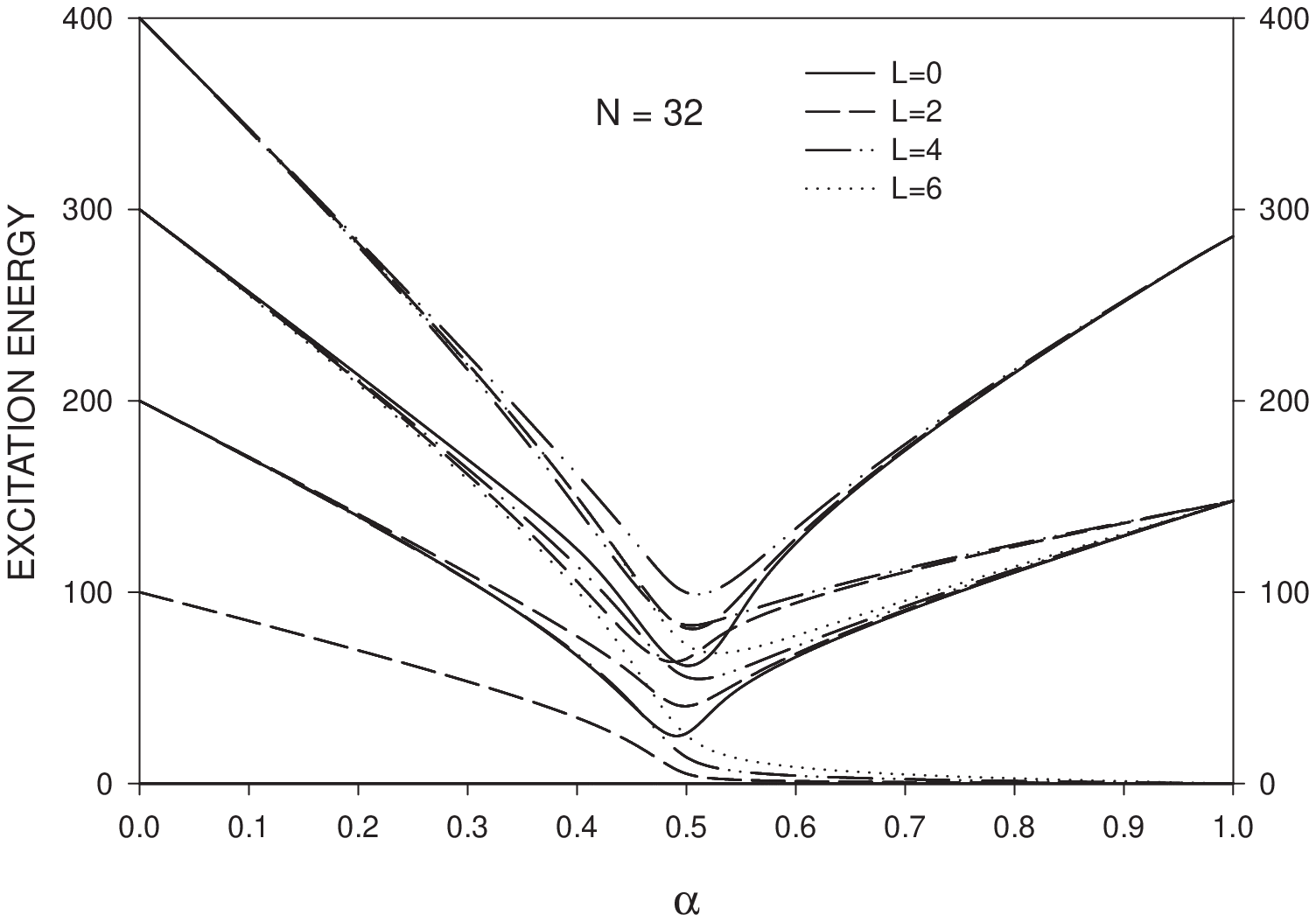}
\hspace{0.1cm}
  \includegraphics[height=.3\textheight,width=0.48\textwidth]{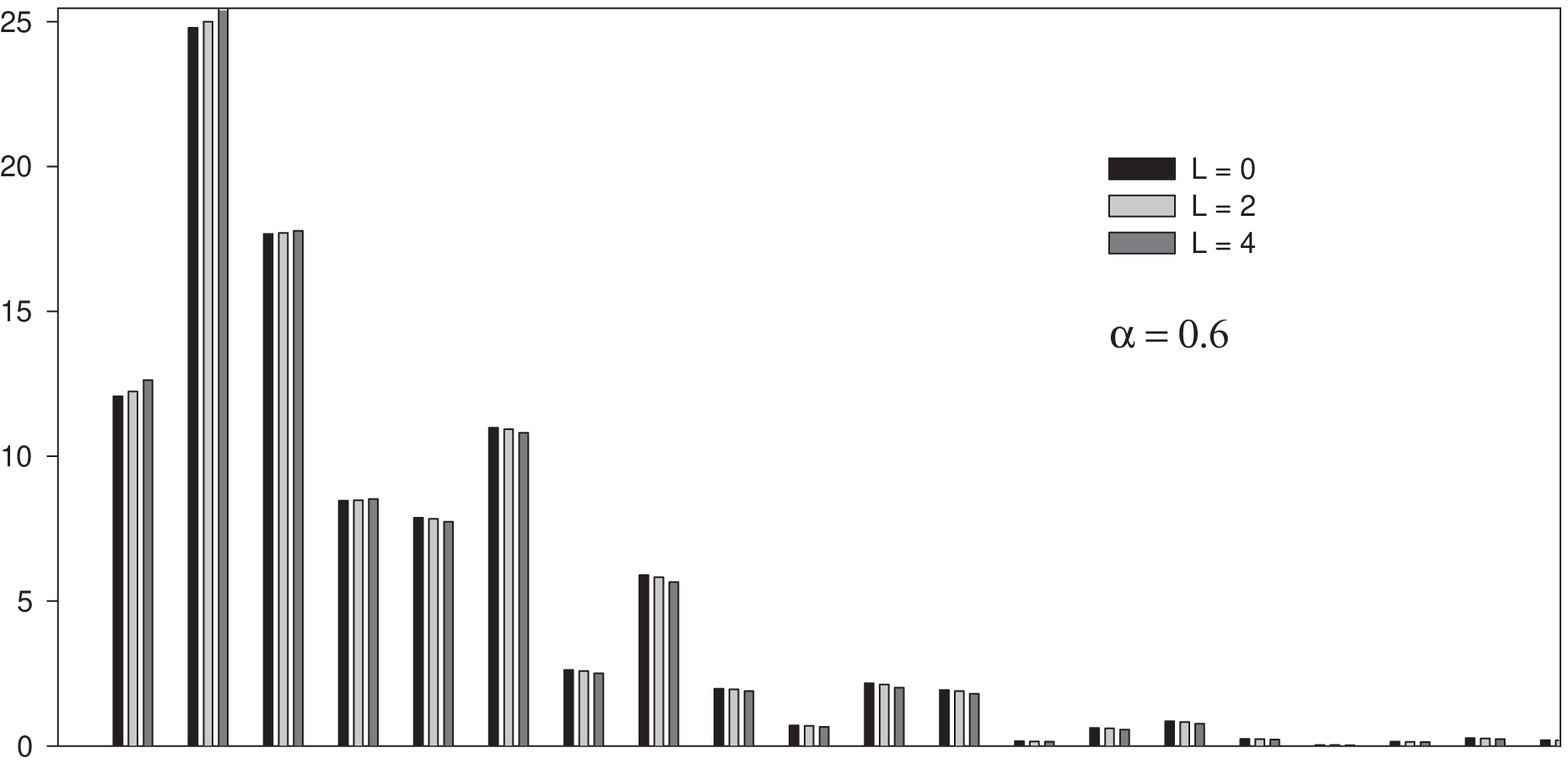}
  \caption{Left panel: energy spectrum for the $U(5)$-$SU(3)$ 
Hamiltonian, Eq.~(4), for $N=32$, as a function of the control 
parameter $\alpha$. 
Right panel: Squared amplitudes for angular momenta $L=0,2,4$ yrast states 
in the $SU(3)$ basis and $\alpha=0.6$. The critical-point value is 
$\alpha_c=0.482$. From~\cite{rowe08}.}
\end{figure}

The IBM can also accommodate spherical to prolate-deformed first-order 
phase transitions, with a high barrier. 
The relevant critical-point Hamiltonian can be transcribed 
in the form~\cite{lev06}
\ba
H(\beta_0) &=& h_{2}\, 
P^{\dagger}_{2}(\beta_0)\cdot\tilde{P}_{2}(\beta_0) ~.
\label{hcri1st}
\ea 
Here $P^{\dagger}_{2\mu}(\beta_0) = 
\beta_{0}\,s^{\dagger}d^{\dagger}_{\mu} + 
\sqrt{7/2}\,\left( d^{\dagger} d^{\dagger}\right )^{(2)}_{\mu}$, 
$\tilde{P}_{2\mu}(\beta_0)=(-1)^{\mu}P_{2,-\mu}(\beta_0)$ 
and $h_2,\,\beta_0>0$. The energy surface of $H(\beta_0)$ coincides with 
the first-order critical surface given in Eq.~(3) and shown in Fig.~(2a). 
For $\beta_0=\sqrt{2}$, $H(\beta_0=\sqrt{2})$ has a subset of 
solvable states with good $SU(3)$ 
symmetry $(\lambda,\mu)=(2N-4k,2k)$~\cite{lev07}, 
where $k=0,1,2,\ldots$
\ba
&&
\vert N,(2N,0)K=0,L\rangle 
\;\; L=0,2,4,\ldots, 2N\qquad E=0~, \qquad\nonumber\\
&&
\vert N,(2N-4k,2k)K=2k,L\rangle
\;\; 
L=K,K+1,K+2,\ldots, (2N-2k)\;\;\;\nonumber\\
&&\qquad 
E=3h_2(2N+1-2k)k\; , \;\; k>0 ~. 
\label{solvsu3}
\ea
For $k=0$, the solvable states are members of a 
prolate-deformed ground band. 
For $k>0$, the solvable states are members of the 
$\gamma^k$ bands, with $K=2k$. 
In addition, $H(\beta_0=\sqrt{2})$ has solvable spherical eigenstates 
with good $U(5)$ symmetry, 
\ba 
&&
\vert N,n_d=\tau=L=0 \rangle
\qquad E = 0~, 
\nonumber\\
&&
\vert N,n_d=\tau=L=3 \rangle 
\qquad E = 3 h_2(2N-1)~.
\ea
The remaining levels in the spectrum, shown 
in Fig.~4, are either predominantly spherical or deformed states 
arranged in several excited bands. 
Their wave functions are spread over many 
$U(5)$ and $SU(3)$ irreps. 
This situation, for which only a subset of states obey 
an exact dynamical symmetry, while other states are mixed, is referred to 
as a partial dynamical symmetry (PDS)~\cite{lev07}. 
For the first-order 
critical-point Hamiltonian considered here, 
some states are solvable with good $U(5)$ symmetry, 
some are solvable with good $SU(3)$ symmetry and all other 
states are mixed with respect to both $U(5)$ and $SU(3)$.  
This behavior defines a $U(5)$ PDS coexisting with a $SU(3)$ PDS.
\begin{figure}[t]
\includegraphics[height=0.3\textheight,width=0.43\textwidth,
angle=270]{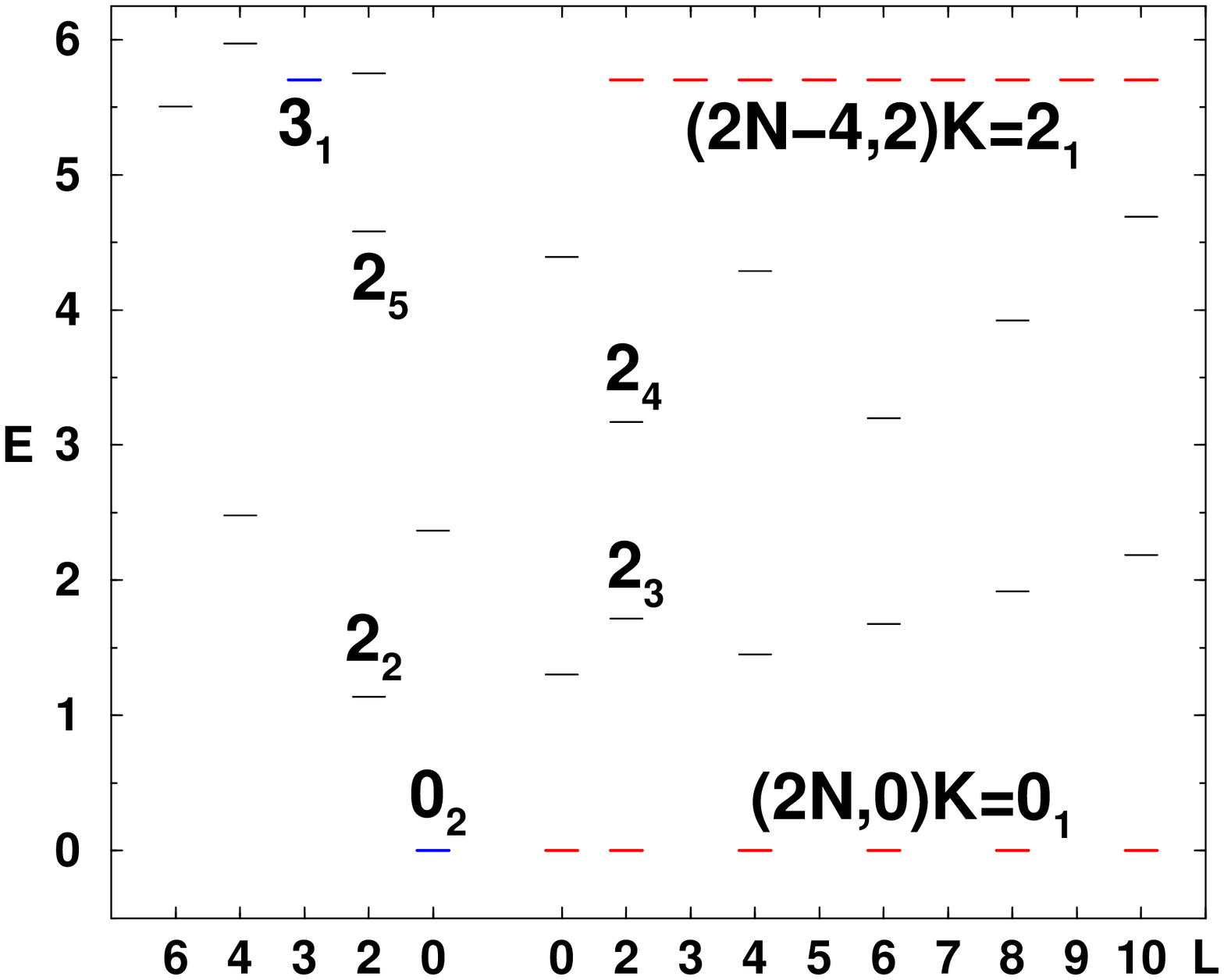}
\hspace{0.3cm}
  \includegraphics[height=0.3\textheight,width=0.43\textwidth,
angle=270]{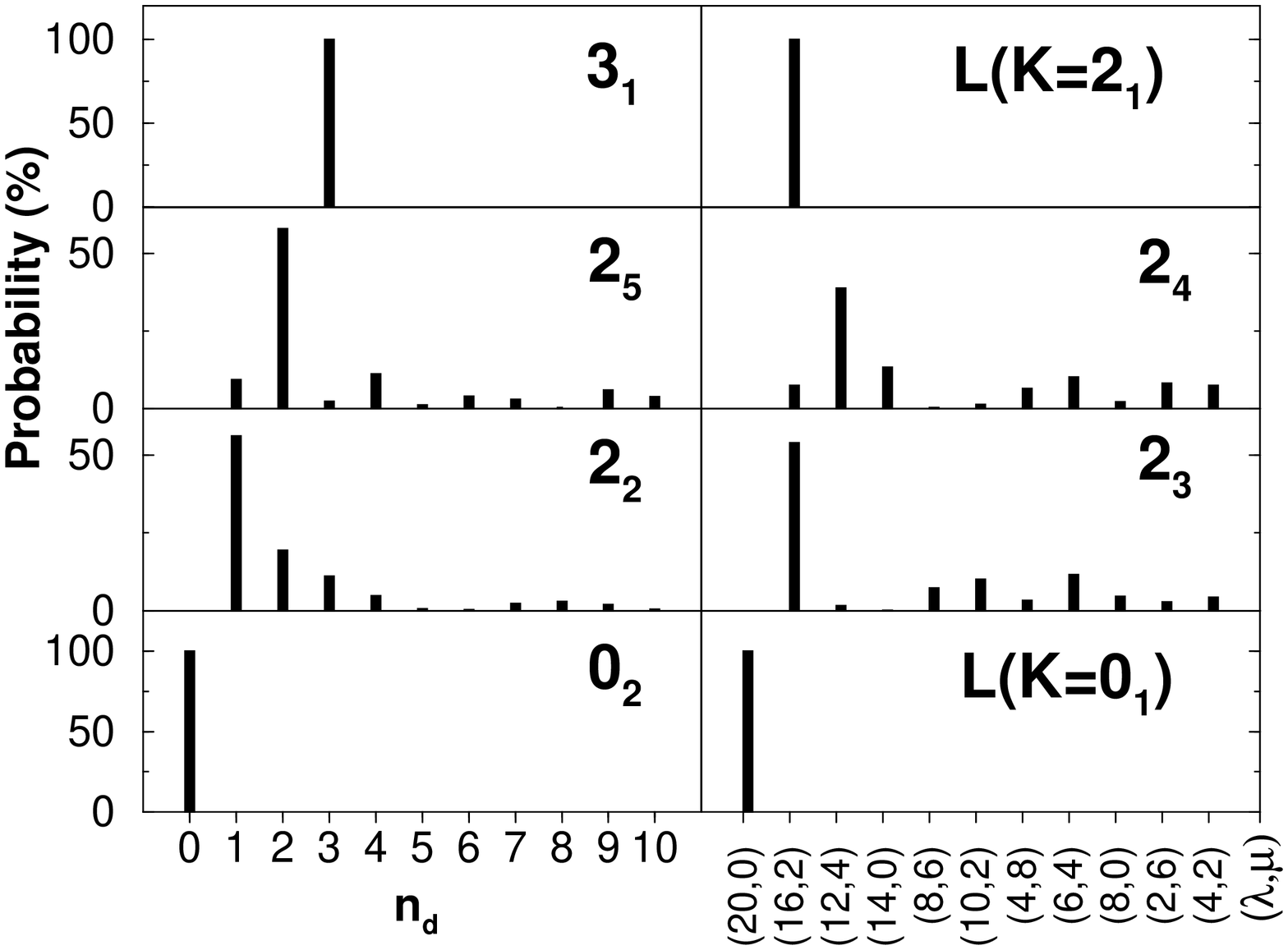}
  \caption{
Left panel: spectrum of $H(\beta_0=\sqrt{2})$, Eq.~(\ref{hcri1st}), 
for $N=10$. 
$L(K=0_1)$ and $L(K=2_1)$ are solvable $SU(3)$ states 
of Eq.~(6) with $k=0,1$ respectively. 
$L=0_2,3_1$ are the solvable $U(5)$ states of Eq.~(7). 
Right Panel: 
$U(5)$ ($n_d$) and $SU(3)$ $[(\lambda,\mu)]$ decomposition of 
selected states. From~\cite{lev07}.}
\end{figure} 

In summary, the study of quantum phase transitions in nuclei provides 
a fertile test-ground for the development of novel concepts of symmetry. 
The latter include ``critical-point symmetries'' and partial dynamical 
symmetries at the critical-point and quasidynamical symmetry away from the 
critical point. 

This work was supported in part by a grant from the U.S.-Israel Binational 
Science Foundation and in part by DOE Grant No. DE-FG-02-91ER40608.

\end{document}